\documentclass{article}
\usepackage[english]{babel}
\usepackage[a4paper,top=2cm,bottom=2cm,left=3cm,right=3cm,marginparwidth=1.75cm]{geometry}

\usepackage{amsmath}
\usepackage{graphicx}
\usepackage[colorlinks=true, allcolors=blue]{hyperref}

\title{CoV-Spectrum: Analysis of globally shared SARS-CoV-2 data to Identify and Characterize New Variants}

\author{Chaoran Chen\,$^{\text{1,2}}$, Sarah Nadeau\,$^{\text{1,2}}$, Michael Yared\,$^{\text{3}}$, Philippe Voinov\,$^{\text{3}}$,\\Ning Xie\,$^{\text{4}}$, Cornelius Roemer\,$^{\text{2,5}}$ and Tanja Stadler\,$^{\text{1,2}}$\\

\footnotesize $^1$ Department of Biosystems Science and Engineering, ETH Zürich, Basel, CH-4058, Switzerland and \\
\footnotesize $^2$ Swiss Institute of Bioinformatics, Lausanne, CH-1015, Switzerland and \\
\footnotesize $^3$ Department of Computer Science, ETH Zürich, Zürich, CH-8092, Switzerland\\
\footnotesize $^4$ Department of Informatics, University of Zurich, Zürich, CH-8050, Switzerland\\
\footnotesize $^5$ Biozentrum, University of Basel, Basel, CH-4056, Switzerland}

\begin{document}
\maketitle

\abstract{\textbf{Summary:} The CoV-Spectrum website supports the identification of new SARS-CoV-2 variants of concern and the tracking of known variants. Its flexible amino acid and nucleotide mutation search allows querying of variants before they are designated by a lineage nomenclature system. The platform brings together SARS-CoV-2 data from different sources and applies analyses. Results include the proportion of different variants over time, their demographic and geographic distributions, common mutations, hospitalization and mortality probabilities, estimates for transmission fitness advantage and insights obtained from wastewater samples.\\
\textbf{Availability and Implementation:} CoV-Spectrum is available at \href{https://cov-spectrum.ethz.ch}{https://cov-spectrum.ethz.ch}. The code is released under the GPL-3.0 license at \href{https://github.com/cevo-public/cov-spectrum-website}{https://github.com/cevo-public/cov-spectrum-website}.}

\section{Introduction}

Most mutations in the SARS-CoV-2 genome do not cause phenotypic changes in the virus. However, some mutations may change the virus such that it is (i) more transmissible, (ii) causes a more severe disease outcome, or (iii) has the ability to evade immunity after infection or vaccination. A new variant with one of these properties is classified as a variant of concern (VOC)~\cite{world_health_organization_special_2021}. 
It is crucial to rapidly identify and characterize new variants of concern such that public health measures can be adapted to emerging threats. 

Demonstrating that that one of the VOC properties (i)-(iii) is met by a new variant in real-time is an ongoing challenge for public health. In particular, observing a quickly spreading variant does not necessarily imply a transmission advantage. In fact, there are many factors that can influence a variant's observed spread: geographic biases and different interventions in different regions, demographic biases and different behaviors in different populations, varying contact tracing efforts, and varying test regimes, among others. For example, the variant named 20E~(EU1) (B.1.177) spread across Europe in summer 2020. However, rather than having a transmission advantage, it appears that superspreading events and travel activities drove this spread~\cite{hodcroftEmergenceSpreadSARSCoV22020}. Therefore, it is important to investigate a broad set of data from different regions to evaluate the risk posed by new variants. This was done for the variant Alpha (B.1.1.7) which showed a consistent relative growth in many countries, indicating an intrinsic  transmission advantage of 43-90$\%$~\cite{Davieseabg3055}.

The CoV-Spectrum website aims to help track known VOCs and facilitate early identification of new ones. It brings together the global public dataset of genomic sequences and additional epidemiological data (section~\ref{sec:data}) to provide a multifaceted view of a variant. The website's variant search feature allow users to track combinations of amino acid and nucleotide mutations, in addition to already designated lineages.

\section{Methods}

\subsection{Data sources and data presentation}\label{sec:data}

The primary data presented by CoV-Spectrum are genomic sequences. We currently provide two instances of CoV-Spectrum: one that uses data provided by GISAID~\cite{elbeGisaid2017}, and another one that uses data from NCBI GenBank provided through Nextstrain~\cite{nextstrain}. These are whole genome sequences of SARS-CoV-2 from countries across the globe as well as basic metadata such as the sampling date, location (often at the level of national divisions) and, for some sequences, the age and sex of the infected individual. We clean the location data with Nextstrain's geo location rules\footnote{\href{https://github.com/nextstrain/ncov-ingest/blob/master/source-data/gisaid_geoLocationRules.tsv}{https://github.com/nextstrain/ncov-ingest/blob/master/source-data/gisaid\_geoLocationRules.tsv}} and we run Nextclade~\cite{aksamentovNextcladeCladeAssignment2021} to obtain aligned nucleotide and amino acid sequences. These data are updated daily.

CoV-Spectrum uses this data to create  plots summarizing the raw data and to perform statistical analyses. The plots include the prevalence, estimated number of cases, demographic and geographic distributions and the common mutations of a variant. Some of the plots can further be stratified by geographic divisions. These are presented in a grid, enabling the user to visually check whether the same dynamic is present in different divisions.

For Switzerland, we receive additional metadata from the Swiss Federal Office of Public Health. The metadata is linked to the whole genome sequences and includes, for example, additional demographic, hospitalization and mortality information. CoV-Spectrum uses this unique dataset to compute the hospitalization and mortality probabilities for different age groups and shows a plot that compares the hospitalization and mortality probabilities of confirmed cases infected with a selected variant with other variants. This enables direct assessment of VOC property (ii) (severe outcome). With this feature, we can see that the hospitalization probability of cases infected with the Alpha variant is indeed higher in older age groups, as suggested by other studies~\cite{Davies2021, Challenn579}.

\subsection{Statistical analysis}\label{sec:analysis}

In addition to presenting the raw data, CoV-Spectrum applies statistical analyses to them. For instance, First, CoV-Spectrum shows the mutations that occur in sequences of a variant and, by ranking them by their Jaccard similarity, it helps identify the mutations that are specific to a particular variant. 

Second, CoV-Spectrum integrates a model to estimate variant transmission fitness advantages, as described in \cite{chenQuantification2021}. \cite{chenQuantification2021} presents static results for the Alpha variant in Switzerland, while CoV-spectrum allows users to explore results for any variants and countries. This enables assessment of VOC property (i) (increased transmissibility). 
For Switzerland, we additionally receive estimates of the proportion of different variants in wastewater samples from collaborators. The underlying procedure is described in~\cite{jahnWastewater2021}, and is currently applied to a selection of variants for which characteristic mutations are manually chosen. This allows to assess if mutations that are identified as spreading in the population based on clinical data is confirmed in wastewater data.

\subsection{Linking other services}

Many COVID-19 dashboards and web tools have been developed since the start of the pandemic, each with their own specific use case. CoV-Spectrum can serve as a hub between several of these services. Namely, the website directly integrates  external services so that users can obtain more information about selected variants at the click of a button. For example, CoV-Spectrum can send a list of sequence identifiers to UShER~\cite{usher}, which will then place the sequences on a pre-defined tree. It can also redirect the user to Taxonium~\cite{taxonium}, which highlights the selected variant in a pre-computed global tree with millions of nodes. Finally, it links to CoVariants~\cite{covariants}, which provides users with curated information about a variant.

\subsection{Sharing of results}\label{sec:sharing}

To promote dissemination of real-time results, CoV-Spectrum's plots and tables are made available to external websites via iframes. These plots remain interactive and will be automatically updated as new data arrives. We used this technique, for example, to integrate plots into a dedicated website explaining the spread of the Alpha variant in Switzerland\footnote{\href{https://cevo-public.github.io/Quantification-of-the-spread-of-a-SARS-CoV-2-variant/}{https://cevo-public.github.io/Quantification-of-the-spread-of-a-SARS-CoV-2-variant/}}.

\subsection{Implementation}

The frontend of CoV-Spectrum is a single-page React application written in TypeScript. It retrieves data from two REST APIs. First, CoV-Spectrum's own server application provides the non-sequence data. Then, our Lightweight API for Sequences (LAPIS)~\cite{lapis} provides the sequence data. LAPIS is a general API to query sequences that is maintained as a separate project. The servers are written in Kotlin and Java using the Spring Boot framework. Finally, the data are stored in a PostgreSQL database.

\section{Conclusion}

The CoV-Spectrum website facilitates rapid detection and characterization of circulating SARS-CoV-2 variants around the globe. The website offers users a convenient way to assess the available SARS-CoV-2 sequencing data together with its metadata. It provides rich information by providing timely figures and tables produced based on globally shared data. As mentioned, evaluating variants requires careful consideration of potential  biases in the sequencing data. CoV-Spectrum aims to help with this task by providing the appropriate geographic and demographic context, wherever possible. Users should, however, be aware of possible sampling biases in the raw data, which may carry through to results presented on CoV-Spectrum. Thus, any results should be interpreted and communicated accordingly.

CoV-spectrum is only possible due to the ongoing efforts of the international community to perform sequencing and make the data rapidly and openly available on GISAID and Genbank. However, some crucial tasks like assessing VOC properties (ii) (severe outcome) and (iii) (immune/vaccine breakthrough) require additional metadata, such as  the severity of infections or vaccine status~\cite{Gomez1043}. We call for global sharing of such metadata. The sharing of properly anonymized and aggregated data will facilitate the rapid identification of VOCs. This will be crucial for timely global public health responses.

\section*{Acknowledgements}
CoV-Spectrum is enabled by data from GISAID. We acknowledge the GISAID team and all data submitters. Further we acknowledge the Federal Office of Public Health in Switzerland for providing metadata for Swiss sequences.

\section*{Funding}
TS acknowledges funding from the Swiss National Science foundation (Special Call on Coronaviruses; 31CA30 196267 and 31CA30 196348).

\bibliographystyle{vancouver}
\bibliography{references.bib}

\end{document}